\documentclass[12pt]{article}
\usepackage{graphics}
\usepackage{epsfig}
\usepackage{subfigure}
\usepackage{graphics}
\usepackage{amsmath}
\usepackage{amsfonts}
\usepackage{amssymb}
\usepackage{url}
\usepackage{hyperref}
\newcommand{\be}{\begin{equation}}
\newcommand{\ee}{\end{equation}}
\newcommand{\bea}{\begin{eqnarray}}
\newcommand{\eea}{\end{eqnarray}}
\newcommand{\bref}[1]{(\ref{#1})}
\begin{document}
\begin{titlepage}
\begin{flushright}
KEK-TH-1246
\end{flushright}
\vspace{4\baselineskip}
\begin{center}
{\Large\bf 
D-term assisted Anomaly Mediation\\
in $E_6$ motivated models
}
\end{center}
\vspace{1cm}
\begin{center}
{\large
Masaki Asano$^{a}$
\footnote{\tt E-mail:masano@icrr.u-tokyo.ac.jp},
Tatsuru Kikuchi$^{b}$
\footnote{\tt E-mail:tatsuru@post.kek.jp}
and Sung-Gi Kim$^{c}$
\footnote{\tt E-mail:sunggi@eken.phys.nagoya-u.ac.jp}
}
\end{center}
\vspace{0.1cm}
\begin{center}
${}^{a}$
{\small \it Institute for Cosmic Ray Research (ICRR),\\
University of Tokyo, Kashiwa, Chiba 277-8582, Japan}\\
${}^{b}$ 
{\small \it Theory Division, KEK,
Oho 1-1, Tsukuba, Ibaraki, 305-0801, Japan}\\
${}^{c}$ {\small \it
Department of Physics, Nagoya University, Nagoya 464-8602, Japan}\\
\medskip
\vskip 5mm
\end{center}
\vskip 5mm
\begin{abstract}
We investigate effects of D-term contributions to the anomaly mediated 
supersymmetry breaking scenario. 
If we introduce an E$_6$ GUT motivated D-term, it is possible
that slepton mass squared is positive at weak scale 
even if there are no other D-term contributions. 
Moreover, as a consequence of additional D-term contributions to scalar masses, 
we obtain various soft supersymmetry breaking mass spectra, which are different
from those obtained in the conventional anomaly mediation scenario.
Then there would be a distinct signature of this scenario at the LHC. For example,
in several cases, there exist mass splittings between the right-handed sfermion and the left-handed sfermion. 
We also discuss some characteristic features of the neutralino dark matter in this model.
\end{abstract}
\end{titlepage}
\section{Introduction}
Supersymmetry (SUSY) extension is one of the most promising way 
 to solve the gauge hierarchy problem in the standard model (SM) \cite{SUSY}.
Since any superpartners have not been observed 
 in current experiments, SUSY should be broken at low energies. 
Furthermore, soft SUSY breaking terms are severely constrained 
 to be almost flavor blind and CP invariant. 
Thus, the SUSY breaking has to be mediated to the visible sector 
not to induce too large CP and flavor violation effects.
Some mechanisms to achieve such SUSY breaking mediation  
 have been proposed \cite{Luty:2005sn}. 

The anomaly mediated supersymmetry breaking (AMSB) scenario
\cite{AMSB1, AMSB2, AMSB3} is one of the most attractive scenario 
 due to its flavor-blindness and ultraviolet (UV) insensitivity 
 for the resultant soft SUSY breaking terms.
The pattern of SUSY breaking does not depend
at all on physics at higher energy scales. 
On the eve of the Large Hadron Collider (LHC) operation at CERN, 
which start this year, there are several studies in the aspects of 
collider physics to discriminate the AMSB scenario from 
the other SUSY breaking mediation \cite{Barr:2002ex, Datta:2002vy, Asai:2007sw}.
Despite the appeal of the AMSB, the original version of the AMSB is excluded. 
The slepton squared masses become negative at the weak scale,
and hence the theory would break U(1)$_{\rm em}$.
There have been many attempts to solve this problem
by incorporating additional positive contributions
to slepton squared masses at tree level \cite{AMSB1, tree1, tree2, tree3, tree4, tree5, tree6, Hodgson:2007kq}
or at quantum level \cite{PR, quantum}.

Among them, adding some D-terms such as ${\rm U}(1)_Y$ and 
${\rm U}(1)_{\rm B-L}$ may be the most interesting possibility \cite{tree5},
because U(1)$_Y$ and U(1)$_{\rm B-L}$ as local symmetries 
are included in SO(10) grand unified theory (GUT) 
\footnote{For the group theoretical aspects of {\rm SO}(10), see for example, \cite{FIKMO}.}.
The idea of GUT's bears several profound features.
Perhaps the most obvious one is that GUT's have the potential to unify 
the diverse set of particle representations and parameters found in 
the SM into a single, comprehensive, and hopefully predictive framework.  
For example, by unifying all U(1) generators within a 
non-Abelian theory, GUT's would provide an explanation for the 
quantization of electric charge.  
By combining GUT's with SUSY, 
we hope to unify the attractive features of GUT's simultaneously 
with those of SUSY into a single theory, SUSY GUT's \cite{Sakai:1981gr}.  
The apparent gauge couplings unification of the 
minimal supersymmetric standard model (MSSM) is strong circumstantial 
evidence in favor of the emergence of a SUSY GUT near 
$M_{\rm GUT} \sim 10^{16}$ [GeV] \cite{Amaldi:1991cn, unification}.
While there are {\it a priori} many choices for possible GUT gauge group,
the list can be narrowed down by requiring groups of rank $\geq 4$ that have 
complex representations. The smallest groups satisfying these requirements are 
SU(5), SU(6), SO(10), and E$_6$.
Among these choices, SO(10) is particularly attractive 
\cite{Fritzsch:1974nn}, because SO(10) is the smallest simple Lie group 
for which a single anomaly-free irreducible representation 
(namely the spinor ${\bf 16}$ representation) 
can accommodate the entire SM fermion content of each generation.
In addition, the recent progress in neutrino physics \cite{Mohapatra:2005wg}
makes SO(10) GUT's \cite{so10} a favorite candidate for GUT's 
because it naturally incorporates the see-saw mechanism 
\cite{see-saw} that can naturally explain the lightness of the light neutrino 
masses.

While it is interesting to add SO(10) motivated D-terms to the AMSB scenario, 
an extension of the grand unification group of SO(10) to E$_6$ is also worthy
of attention \cite{E6, Bando:1999km, tree3}.
In the E$_6$ GUT, the fundamental representation ${\bf 27}$ includes 
${\bf 10}$ as well as ${\bf 16}$ of SO(10) in the following manner:
\bea
{\bf 27} &\to& {\bf 16}_{+1} + {\bf 10}_{-2} + {\bf 1}_{+4}
\nonumber\\
&\to& [{\bf 10}_{-1} + \overline{\bf 5}_{+3} + {\bf 1}_{-5} ] + [{\bf 5}_{+2} + \overline{\bf 5}_{-2} ] + {\bf 1}_0
\label{decom}
\eea
under $E_6 \supset {\rm SO}(10) \times {\rm U}(1)_V,~{\rm SO}(10) \supset {\rm SU}(5) \times {\rm U}(1)_W$.
Hereafter we take the following linear combination of those two U(1) charges 
to match with the convention in \cite{Martin}.
\be
S = -(5V+W)/12,~~X=W/3 \;.
\ee
The symmetry breaking pattern of SO(10) down to the SM gauge group can be found in the literature \cite{FIKMO}. 
A linear combination of U(1)$_X$ charge and U(1)$_{\rm B-L}$ charge gives a correct SM hypercharge normalization
under the following relation (see, for example, \cite{FIKMO}).
\be
Y = \frac{1}{4} \left[3 X + 5 (B-L) \right] \;.
\ee
While there are several studies on the extra gauge boson, $Z^\prime$, 
which is associated with an extra U(1) gauge symmetry, potentially arisen from E$_6$ GUT,
the effect of the U(1)$^\prime$ D-term has not been studied well in the literature.
Hereafter, we label the U(1)$^\prime$ gauge symmetry which is originated from E$_6$ GUT
as U(1)$_S$.
In this paper, we specifically consider the effect of such an extra U(1)$_S$ D-term,
in addition to the ${\rm U}(1)_Y$ and the ${\rm U}(1)_{\rm B-L}$ D-terms
to the soft SUSY breaking mass terms.
Since the effects of this extra U(1)$_S$ D-term to the sfermion mass squared gives
common signature contributions, the tachyonic slepton problem can simply be solved
in this framework.

This paper is organized as follows. In section 2, we begin with a brief review of AMSB
and explain a model of D-term assisted anomaly mediation in E$_6$ GUT
motivated models.
In section 3, we show numerical results for the calculation of sparticle mass spectra and 
a set of allowed parameter spaces of $(D_Y, D_{\rm B-L}, D_{\rm S})$.
After calculating all the sparticle masses by using {\tt ISAJET 7.75} \cite{ISAJET},
the relic density of the cold dark matter (CDM) in this scenario is also estimated by using
{\tt micrOMEGAs} \cite{Belanger:2006is}.
The last section is devoted to summary and discussions.

\section{D-term assisted anomaly mediation}
In this section, we work out in the superconformal framework 
 of supergravity \cite{Kaku:1978nz}, 
and we explain the D-term assisted anomaly mediation scenario.

In the superconformal framework of supergravity, 
the basic Lagrangian is given by 
\bea
 {\cal L}_{\rm SUGRA} &=& - 3 \int d^4 \theta \;\phi^\dag \phi \; e^{-K/3}  
  + \int d^2 \theta \;\phi^3 W + h.c. \; , 
\eea
where $\phi=1+ \theta^2 F_{\phi}$ is the compensating multiplet,
$K$ is the K\"ahler potential in the conformal frame, $W$ is the superpotential,
and the reduced Planck mass is set to unity.

As for the gauge sector in the MSSM, the kinetic term is of the form, 
\bea
{\cal L}_{\rm gauge} = \frac{1}{4} \int d^2 \theta \;
 \tau_a\left(\frac{\mu_R}{\Lambda \phi} \right)  {\cal W}^{a \alpha} {\cal W}^a_\alpha \; . 
\label{gaugeK}
\eea
At the classical level, the compensator $\phi$ does not appear
in the gauge kinetic term as the gauge chiral superfield ${\cal W}^{a \alpha}$
has a chiral weight $\frac{3}{2}$. 
It turns out that the dependence of $\phi$ comes out radiatively
through the cutoff scale $\Lambda$ ($\mu_R$ is the renormalization scale).
In the above setup, non-zero $F_\phi$ induces soft SUSY breaking terms 
 through the AMSB, and the resultant SUSY breaking mass scale is 
 characterized by $m_{\rm AMSB} \sim F_\phi/(16 \pi^2) \equiv M_{\rm SUSY}$. 
Considering the anomaly mediation contribution to the soft scalar masses and A-terms, 
we take the minimal K\"ahler potential for the MSSM superfields, 
$K_{\rm MSSM} = Q_i^\dagger e^{2g_a V_a} Q_i$, where $Q_i$ stands for the MSSM
matter and Higgs superfields. 
Expanding $e^{K/3}$, the K\"ahler potential for the MSSM superfields is described as 
\be 
{\cal L}_{\rm kin} = \int d^4 \theta \,\phi^\dagger \phi \,
 Q_i^\dagger e^{2g_a V_a} Q_i  + \cdots \; .
\label{matterK}
\ee
As discussed in Ref. \cite{method}, in softly broken supersymmetry, the soft terms associated to a chiral
superfield $Q_i$ can be collected in a running superfield wave function ${\cal Z}_i(\mu_R)$ such that
\be
\ln {\cal Z}_i(\mu_R) = \ln Z_i(\mu_R) + [A_i(\mu_R) \theta^2 + h.c.] - \widetilde{m}_i^2 (\mu_R) \theta^4 \;.
\ee
The running wave functions can be defined as $Z_i(\mu_R) = c_i(p^2 = -\mu_R^2)$, 
where $c_i$ is the coefficient of $Q_i^\dag Q_i$ in the one point-irreducible (1PI) effective action.
Therefore, turning on superconformal anomaly amounts to the shift $\mu_R \to \mu_R/(\phi^\dag \phi)^{1/2}$.
\be
{\cal Z}_i(\mu_R) = Z_i\left(\frac{\mu_R}{(\phi^\dag \phi)^{1/2}}\right)\;.
\ee
According to the method developed in Ref.~\cite{method} (see also Ref.~\cite{PR}), 
soft SUSY breaking terms (each gaugino masses $M_a$, sfermion squared masses
$\widetilde{m}_i^2$ and $A$-parameters) can be extracted from renormalized
gauge kinetic functions and SUSY wave function renormalization coefficients, 
\bea 
&& M_{a}(\mu_R) = b_a g_a^2(\mu_R) M_{\rm SUSY} \; , 
 \nonumber\\ 
&& \widetilde{m}_i^2(\mu_R) = - 8 \pi^2 \mu_R   \frac{d\gamma_i (\mu_R)}{d \mu_R}  M_{\rm SUSY}^2 \; , 
 \nonumber \\
&& A_{ijk}(\mu_R) =- \left[ 
   \gamma_i (\mu_R) +\gamma_j (\mu_R) + \gamma_k (\mu_R) \right] M_{\rm SUSY} \; .
\label{softterms}
\eea
Here, $g_a$ are the gauge couplings, 
      $b_a$ are beta function coefficients, and 
   $\gamma_i$ are anomalous dimensions of the MSSM
matter and Higgs superfields.
The results in Eq.~(\ref{softterms}) are true at any energy scale,
and all the soft mass parameters can be described by only one parameter, 
$F_\phi$, so the anomaly mediation is highly predictive.
This indicates that the soft terms at a low-energy scale depend only on anomalous dimensions 
or beta functions at that scale and do not care about the theory at higher energies. 
This UV insensitivity is the main feature of the anomaly mediation.

There are remaining two parameters in the Higgs sector, 
 namely $\mu$ and $B \mu$ terms, 
 that are responsible for electroweak symmetry breaking
 and should be of the order of the electroweak scale. 
 The natural value of the $B$-parameter would be 
 $B \sim F_{\phi} \gg M_{\rm SUSY}$, and 
 the Higgs sector should be extended in order to 
 achieve the $B$-parameter being at the electroweak scale. 
Although some mechanism is required to realize $\mu \sim B \sim M_{\rm SUSY}$,
in the following analysis, we treat them as free parameters.
That is, $\mu$ and $B \mu$ are replaced into two free parameters $\tan \beta$ and ${\rm sgn}(\mu)$, 
while the value of $|\mu|$ is determined by the stationary condition of the Higgs potential. 

In the following, we consider to add three D-terms of ${\rm U}(1)_{Y}$, ${\rm U}(1)_{\rm B-L}$ 
and ${\rm U}(1)_S$, and hence total set of free parameters in our analysis is 
$(\tan \beta,~{\rm sgn}(\mu),~M_{\rm SUSY},~D_{\rm Y},~D_{\rm B-L},~D_{\rm S})$.

Now we turn to the discussion to introduce the D-terms to the anomaly mediation. 
 If there exists an extra $U(1)$ gauge multiplet, $V$, having a non-zero $D$-term, 
$\left<V \right> = \theta^2 \bar{\theta}^2 D$,
 the kinetic term of a matter superfield gives
\be
{\cal L} = \int d^4 \theta \, Q_i^\dag e^{q_i V} Q_i
 \supset q_i D\, \widetilde{Q}_i^\dagger 
 \widetilde{Q}_i \;,
\ee
where $q_i$ is the ${\rm U}(1)$ charge of the chiral multiplet $Q_i$. 
This leads to a shift for the scalar squared mass, 
\bea
\widetilde{m}_i^2 \to \widetilde{m}_i^2 - q_i D \;.
\eea
The D-term shifts in the soft masses within the framework of minimal supergravity 
(mSUGRA) has firstly been done in \cite{Martin}.

The ${\rm U}(1)$ symmetry providing the D-term should be anomaly-free. 
As such a ${\rm U}(1)$ symmetry, there exist two candidates in the MSSM, 
 namely ${\rm U}(1)_{Y}$ and gauged ${\rm U}(1)_{\rm B-L}$.
In addition to them, we introduce ${\rm U}(1)_S$ gauge symmetry motivated by
 ${\rm E}_6$ grand unified theories.
Introduction of both ${\rm U}(1)_{\rm B-L}$ and ${\rm U}(1)_S$ gauge symmetries 
 is indeed well-motivated, if we assume that the MSSM is embedded into 
 a GUT based on a higher rank gauge group such as ${\rm E}_6$
 which includes the gauged ${\rm U}(1)_{\rm B-L}$ and ${\rm U}(1)_S$ as a subgroup. 
This possibility is our motivation to consider the D-terms in addition to the anomaly mediation. 
Once we introduce such D-terms for either ${\rm U}(1)_{\rm B-L}$ or ${\rm U}(1)_S$,
non-zero D-term for hypercharge ${\rm U}(1)_{Y}$ will, in general, be induced
through the kinetic mixing
\be
{\cal L} = \lambda \int d^2 \theta \,{\cal W}^\alpha_{\rm B-L} {\cal W}^\alpha_Y
= \xi D_Y \;,
\label{kin}
\ee
where $\xi  = \lambda D_{\rm B-L}$.

Normally, many extra Higgs fields are involved in such models, 
 and some of them have non-zero vacuum expectation values 
 to break the GUT symmetry at the supersymmetric level. 
Once soft SUSY breaking terms for these Higgs fields 
 are included, 
 the vacuum would be realized at the point 
 slightly away from the D-flat directions, 
 so that non-zero D-terms are developed. 
Although it depends on the detailed structure 
 of the Higgs sector, 
 we may naturally expect the scale of the D-term 
 to be $D \sim M_{\rm SUSY}^2$. 
 
The D-terms contributions change sfermion mass spectrum 
 from the one in the conventional anomaly mediation scenario. 
As a result, the sparticle mass spectrum in our scenario 
 can be quite different from the one obtained 
 in the pure anomaly mediation scenario.
\begin{table}
\begin{center}
\begin{tabular}{|c|c|c|c|c|c|c|c|}
\hline \hline
 & $Q$ & $u$ & $d$ & $L$ & $e$ & $H_2$ & $H_1$ \\
\hline
$U(1)_Y$ & ${1/6}$ & ${-2/3}$ & ${1/3}$ & ${-1/2}$ & $1$ & ${1/2}$ & ${-1/2}$\\
\hline
$U(1)_{\rm B-L}$ & ${1/3}$ & ${-1/3}$ & ${-1/3}$ & ${-1}$ & ${1}$ & ${0}$ & ${0}$\\
\hline
$U(1)_{\rm S}$ & ${-1/3}$ & ${-1/3}$ & ${-2/3}$ & ${-2/3}$ & ${-1/3}$ & ${2/3}$ & ${1}$\\
\hline \hline
\end{tabular}
\caption{$U(1)$ charges of MSSM chiral superfields}
\end{center}
\end{table}
Calculating the anomalous dimensions and
taking D-term contributions from ${\rm U}(1)_Y$, ${\rm U}(1)_{\rm B-L}$, and ${\rm U}(1)_S$ 
into account, the soft scalar masses for the first two generations 
are explicitly written as\footnote{
There are some intriguing E$_6$ SUSY GUT models which employ
`E-twisting' to produce variety of hierarchical structures of quark and 
lepton Yukawa couplings in a simple way \cite{Bando:1999km}.
However, in the following, we simply assume that all the MSSM matter and Higgs fields are contained in 
${\bf 16}$ and ${\bf 10}$ of Eq.~(\ref{decom}) respectively.
It might be unclear to extend SO(10) to E$_6$, but an additional U(1) gauge symmetry included in E$_6$
can provide a D-term that may simply solve the tachyonic slepton problem.
In that sense, E$_6$ extension is motivated.
}:
\bea
m_{\widetilde{q}_{1,2}}^2 &=&M_{\rm SUSY}^2
\left[8 g_3^4 - \frac{3}{2} g_2^4 - \frac{11}{18} g_Y^4
-\frac{1}{6} \alpha_{Y} - \frac{1}{3} \alpha_{\rm B-L} + \frac{1}{3} \alpha_{\rm S} 
\right] \;,
\nonumber\\
m_{\widetilde{u}_{1,2}}^2 &=&M_{\rm SUSY}^2
\left[8 g_3^4 - \frac{88}{9} g_Y^4
+ \frac{2}{3} \alpha_{Y} + \frac{1}{3} \alpha_{\rm B-L} + \frac{1}{3} \alpha_{\rm S} 
\right] \;,
\nonumber\\
m_{\widetilde{d}_{1,2}}^2 &=& M_{\rm SUSY}^2
\left[8 g_3^4 - \frac{22}{9} g_Y^4
- \frac{1}{3} \alpha_{Y} + \frac{1}{3} \alpha_{\rm B-L} + \frac{2}{3} \alpha_{\rm S}
\right] \;,
\nonumber\\
m_{\widetilde{\ell}_{1,2}}^2 &=&M_{\rm SUSY}^2 
\left[-\frac{3}{2} g_2^4 - \frac{11}{2} g_Y^4
+ \frac{1}{2} \alpha_{Y} + \alpha_{\rm B-L} + \frac{2}{3} \alpha_{\rm S}
\right] \;,
\nonumber\\
m_{\widetilde{e}_{1,2}}^2 &=& M_{\rm SUSY}^2 
\left[- 22 g_Y^4
- \alpha_{Y} - \alpha_{\rm B-L} + \frac{1}{3} \alpha_{\rm S}
\right]\;.
\eea
Here $g_Y$ is the $U(1)_Y$ gauge coupling constant and is related to the GUT 
normalized one as $g_1^2 = (5/3) g_Y^2$, and we have defined 
$\alpha_Y$, $\alpha_{\rm B-L}$, and $\alpha_{\rm S}$ as 
\bea
\alpha_Y &\equiv& \frac{D_Y}{M_{\rm SUSY}^2} \; ,
\nonumber\\
\alpha_{\rm B-L} &\equiv& \frac{D_{\rm B-L}}{M_{\rm SUSY}^2} \; ,
\nonumber\\
\alpha_{\rm S} &\equiv& \frac{D_{\rm S}}{M_{\rm SUSY}^2} \; ,
\eea
and Yukawa couplings of the first two generations 
 have been neglected as a good approximation. 
Since the D-term contributions are determined by the corresponding U(1) charge,
we list up the charges of $U(1)_Y$, $U(1)_{\rm B-L}$, and $U(1)_{\rm S}$ in Table 1.

For the third generation sfermion masses, Yukawa couplings are involved; 
\bea
m_{\widetilde{q}_{3}}^2 &=& M_{\rm SUSY}^2
\left[8 g_3^4 - \frac{3}{2} g_2^4 - \frac{11}{18} g_Y^4 + y_t^2 b_{y_t} + y_b^2 b_{y_b}
- \frac{1}{6} \alpha_{Y} - \frac{1}{3} \alpha_{\rm B-L} 
+ \frac{1}{3} \alpha_{\rm S} 
\right] \;,
\nonumber\\
m_{\widetilde{u}_3}^2 &=& M_{\rm SUSY}^2
\left[8 g_3^4 - \frac{88}{9} g_Y^4 + 2 y_t^2 b_{y_t} 
+ \frac{2}{3} \alpha_{Y} + \frac{1}{3} \alpha_{\rm B-L} 
+ \frac{1}{3} \alpha_{\rm S}
\right] \;,
\nonumber\\
m_{\widetilde{d}_3}^2 &=& M_{\rm SUSY}^2
\left[8 g_3^4 - \frac{22}{9} g_Y^4  + 2 y_b^2 b_{y_b}
- \frac{1}{3} \alpha_{Y} + \frac{1}{3} \alpha_{\rm B-L} + \frac{2}{3} \alpha_{\rm S}
\right]\;,
\nonumber\\
m_{\widetilde{\ell}_{3}}^2 &=& M_{\rm SUSY}^2
\left[-\frac{3}{2} g_2^4 - \frac{11}{2} g_Y^4  + y_\tau^2 b_{y_\tau}  
+ \frac{1}{2} \alpha_{Y} + \alpha_{\rm B-L} + \frac{2}{3} \alpha_{\rm S}
\right] \;,
\nonumber\\
m_{\widetilde{e}_3}^2 &=&M_{\rm SUSY}^2
\left[- 22 g_Y^4 + 2 y_\tau^2 b_{y_\tau} 
- \alpha_{Y} - \alpha_{\rm B-L} + \frac{1}{3} \alpha_{\rm S}
\right] \;,
\eea
where $b_{y_t}$, $b_{y_b}$ and $b_{y_\tau}$ are given by 
\bea
b_{y_t} &=& 6 y_t^2 + y_b^2 
- \frac{16}{3} g_3^2 - 3g_2^2 - \frac{13}{9} g_Y^2 \;,
\nonumber\\
b_{y_b} &=& y_t^2 + 6 y_b^2 + y_\tau^2 
- \frac{16}{3} g_3^2 - 3g_2^2 - \frac{7}{9} g_Y^2 \;,
\nonumber\\
b_{y_\tau} &=& 3 y_b^2 + 4 y_\tau^2 
- 3g_2^2 - 3 g_Y^2 \;.
\eea
Also, the Higgs soft masses are given by
\bea
m_{H_1}^2 &=& M_{\rm SUSY}^2
\left[-\frac{3}{2} g_2^4 - \frac{11}{2} g_Y^4 + 3 y_b^2 b_{y_b} + y_\tau^2 b_{y_\tau} 
+ \frac{1}{2} \alpha_{Y} - \alpha_{S}
 \right] \;,
\nonumber\\
m_{H_2}^2 &=& M_{\rm SUSY}^2
\left[-\frac{3}{2} g_2^4 - \frac{11}{2} g_Y^4 + 3 y_t^2 b_{y_t} 
- \frac{1}{2} \alpha_{Y} - \frac{2}{3} \alpha_{S}
\right] \;.
\eea
The Higgs mass parameters, $\mu$-term and $B \mu$-term, 
 are determined by the electroweak symmetry breaking conditions, 
\bea
|\mu|^2 &=& 
\frac{m_{H_1}^2 - m_{H_2}^2 \tan^2 \beta}{\tan^2 \beta -1}
- \frac{1}{2} M_Z^2  \;,
 \nonumber\\
B \mu &=& \frac{1}{2} 
 \left[m_{H_1}^2 + m_{H_2}^2  + 2 |\mu|^2 \right] \sin 2 \beta \; .
\label{mu}
\eea
The $A$-parameters in the AMSB scenario are given by
\bea
 A_{ijk} = - \left( \gamma_i +\gamma_j + \gamma_k \right)M_{\rm SUSY} \; 
\eea 
 with the following anomalous dimensions, 
\bea
 \gamma_{q_i} &=& -\frac{8}{3} g_3^2 - \frac{3}{2} g_2^2 - \frac{1}{18} g_Y^2
-( y_t^2+y_b^2) \delta_{i 3} \;, \nonumber \\
 \gamma_{u^c_i}&=&  -\frac{8}{3} g_3^2 - \frac{8}{9} g_Y^2
- 2 y_t^2  \delta_{i 3} \; ,  \nonumber \\
 \gamma_{d^c_i} &=& -\frac{8}{3} g_3^2 - \frac{2}{9} g_Y^2
- 2 y_b^2  \delta_{i 3} \; ,  \nonumber \\
 \gamma_{\ell_i} &=& -\frac{3}{2} g_2^2 - \frac{1}{2} g_Y^2
-  y_\tau^2  \delta_{i 3} \; ,  \nonumber \\
 \gamma_{e^c_i} &=& - 2 g_Y^2
- 2 y_\tau^2 \delta_{i 3} \; ,  \nonumber \\
 \gamma_{H_1} &=& -\frac{3}{2} g_2^2 - \frac{1}{2} g_Y^2
- 3 y_b^2 -y_\tau^2  \; , \nonumber \\
 \gamma_{H_2} &=& -\frac{3}{2} g_2^2 - \frac{1}{2} g_Y^2
- 3 y_t^2  \; .
\eea
Finally, the gaugino masses are given by 
\bea
M_1 &=&  11 g_Y^2 M_{\rm SUSY} \; , 
 \nonumber\\
M_2 &=& g_2^2 M_{\rm SUSY} \; ,
 \nonumber\\
M_3 &=& - 3 g_3^2 M_{\rm SUSY} \; .
\eea
Now we will see the condition to avoid the tachyonic sfermions.
First, once we assume $\alpha_{\rm S} = 0$ as in the SO(10) models,
the condition to have the slepton squared masses to be positive is written by
\be
\alpha_Y < -\alpha_{\rm B-L} < \frac{1}{2}\, \alpha_Y ~~ (\mbox{if}~~\alpha_{\rm S} = 0) \; ,
\label{ineq}
\ee
which can be satisfied with $\alpha_Y <0$ and $\alpha_{\rm B-L}>0$. 
%
If we take $\alpha_Y <0$ to avoid the tachyonic sleptons, it gives positive
contribution to the up-type Higgs soft mass squared. So, too large $\alpha_Y (<0)$ can not
lead to the correct electroweak symmetry breaking.
%
Therefore, the parameter space of ($\alpha_Y,~\alpha_{\rm B-L}$)  is very restricted by these constraints.

In contrast, there is a remarkable effect by adding a non-zero $\alpha_{\rm S}$ term.
The D-term contribution from U(1)$_S$ gives the same sign contribution for
all the soft scalar masses of squarks and sleptons.
Hence, it is indeed helpful to solve the tachyonic slepton problem in the AMSB scenario.
Specifically, one can cure the tachyonic slepton problem by considering only non-zero $D_{\rm S}$ term
without including either $D_{B-L}$ or $D_Y$ term.
As we have seen in Eq.~\bref{kin}, once we introduce several U(1) gauge symmetries at the same time,
in general, one can write down the kinetic mixing between those two U(1) gauge fields.
Then if one of the D-terms has non-zero value, the other D-term would also have non-zero value.
However, if one consider the situation where several U(1) gauge symmetries do not exist 
at the same time, then it is possible to have only one non-zero D-term.
As such an example, we can take the following gauge symmetry breaking,
$E_6 \to {\rm SO}(10) \times {\rm U}(1) \to {\rm MSSM}$,
to consider only one D-term contribution, $D_{\rm S}$ term.

%

\section{Numerical results}
\subsection{SUSY mass spectrum\label{spectrum}}
In this section, we evaluate the sparticles mass spectra
by using {\tt ISAJET 7.75} \cite{ISAJET} in Fig~\ref{Fig2}, Fig.~\ref{Fig3} and
in Table.~\ref{table1}.
Our model includes the parameter set ($\alpha_Y$, $\alpha_{\rm B-L}$, $\alpha_{\rm S}$)
which is introduced in the previous section, the typical soft SUSY breaking mass
scale, $M_{\rm SUSY}$, and $\tan \beta$.
We input all these parameters at the GUT scale, and evolve down to the weak scale
according to the MSSM RGEs
by assuming all the additional U(1) symmetry breaking scale to be at the GUT scale.

\begin{figure}[p]
\begin{center}
\subfigure[$(\alpha_Y,~\alpha_{\rm B-L})$ with a given $\alpha_S= 0$.]
{\includegraphics[width=.4\linewidth]{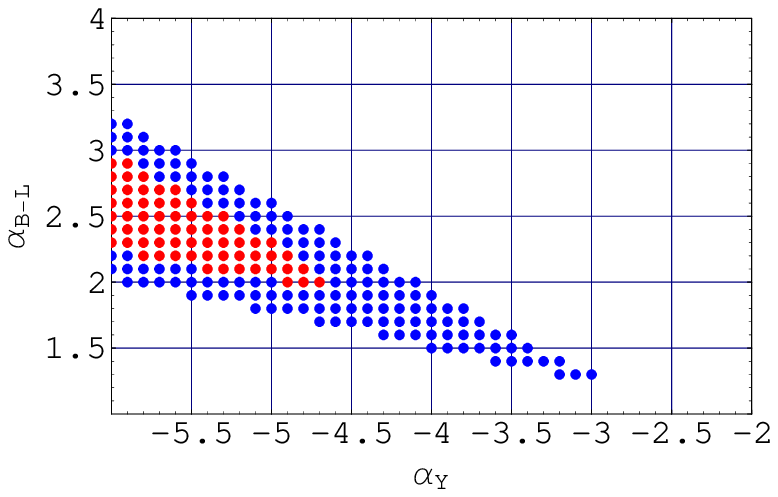}\label{Fig1a}}
\subfigure[$(\alpha_Y,~\alpha_{\rm B-L})$ with a given $\alpha_S= 1$.]
{\includegraphics[width=.4\linewidth]{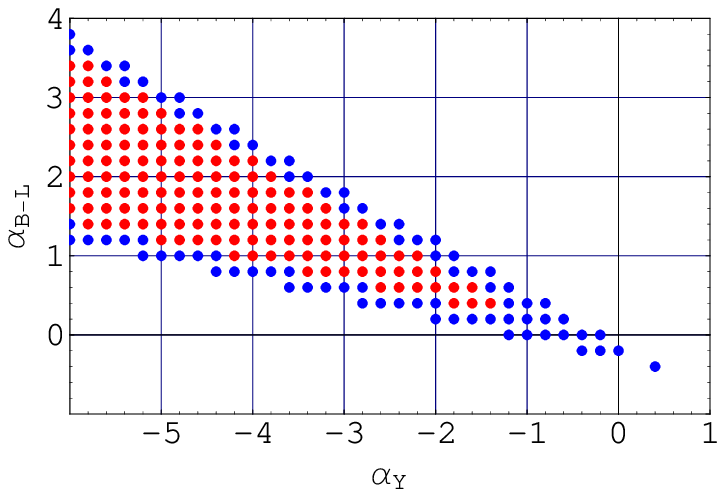}\label{Fig1b}}
\subfigure[$(\alpha_Y,~\alpha_{\rm B-L})$ with a given $\alpha_S= 2$.]
{\includegraphics[width=.4\linewidth]{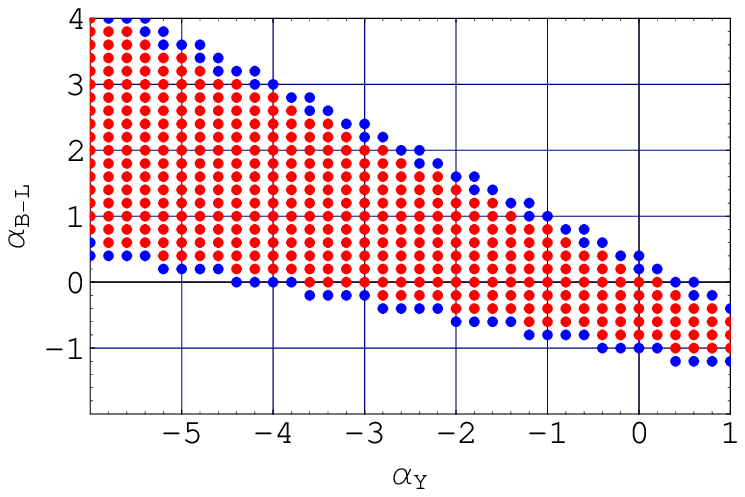}\label{Fig1c}}
\subfigure[$(\alpha_Y,~\alpha_{\rm B-L})$ with a given $\alpha_S= 4$.]
{\includegraphics[width=.4\linewidth]{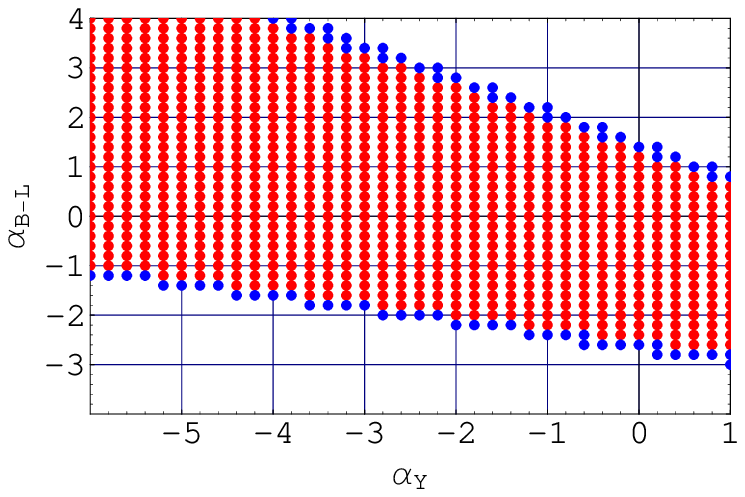}\label{Fig1d}}
\end{center}
\caption{
The allowed parameter set $(\alpha_Y,~\alpha_{\rm B-L})$ with fixed $\alpha_S= 0,\,1,\,2,\,4$,
 which provides all the sfermion squared masses positive  
 and the correct electroweak symmetry breaking 
 in the case of $\tan \beta=10$ and $M_{\rm SUSY}=500$ GeV.
Red colored region is the region where the neutralino becomes LSP,
while the blue colored region is the region where the stau becomes LSP.}
\label{Fig1}
\vspace{1cm}
\end{figure}

\begin{figure}[p]
\begin{center}
\includegraphics[height=.42\linewidth]{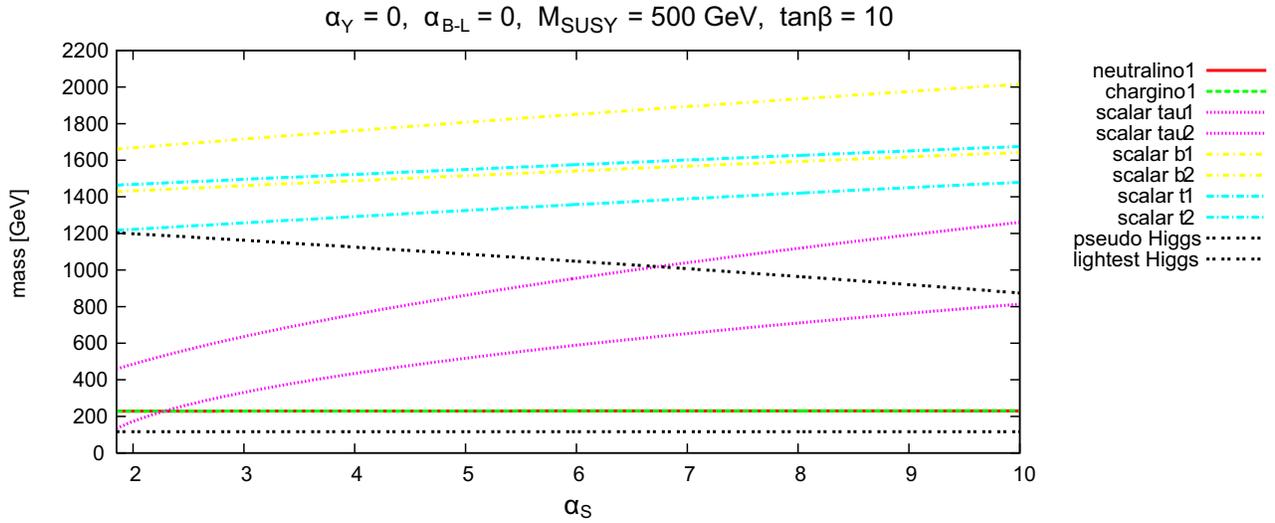}
\end{center}
\caption{
The sparticle and Higgs boson mass spectra (in units of GeV) 
as a function of $\alpha_{\rm S}$ with a fixed set of D-terms as
$(\alpha_Y,\,\alpha_{\rm B-L})=(0,\,0)$, which provides 
all the sfermion squared masses positive 
and the correct electroweak symmetry breaking.
Here, we took $\tan \beta=10$ and $M_{\rm SUSY} = 500$ GeV.
}
\label{Fig2}
\vspace{1cm}
\end{figure}

\begin{figure}[p]
\begin{center}
\includegraphics[height=.42\linewidth]{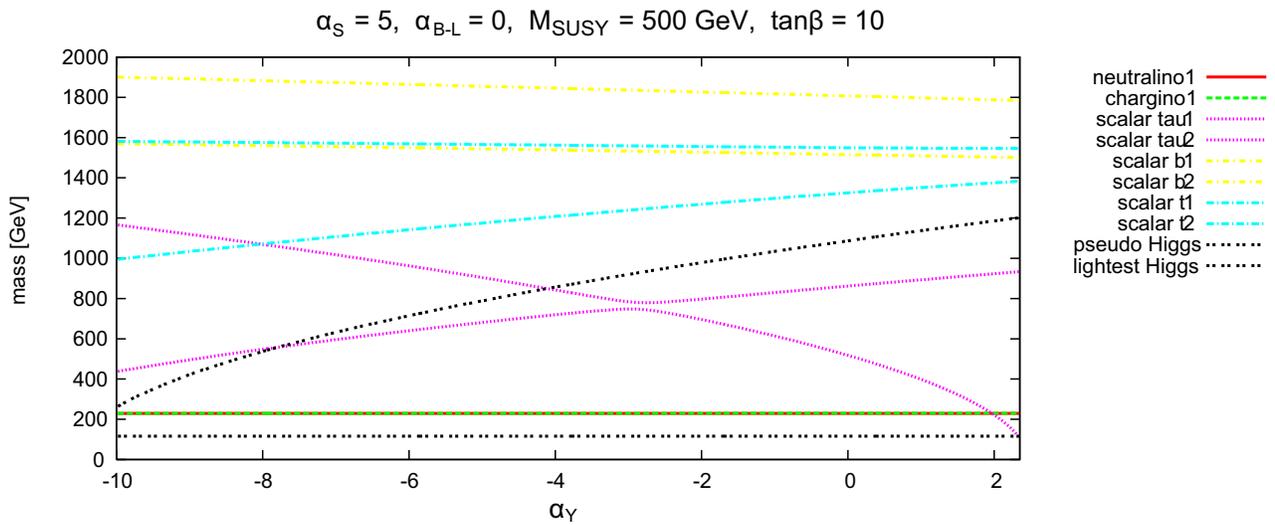}
\end{center}
\caption{
The same figure as Fig.~\ref{Fig2} but as a function of $\alpha_Y$
with a fixed set of D-terms as $(\alpha_{\rm B-L}, \, \alpha_S)=(0,\,5)$.
}
\label{Fig3}
\vspace{1cm}
\end{figure}

First we examine the allowed region of the parameter space 
 ($\alpha_Y$, $\alpha_{\rm B-L}$, $\alpha_{\rm S}$) 
 for given $\tan \beta=10$ and $M_{\rm SUSY}=500$ GeV. 
Sparticle mass spectrum for various inputs in the range of 
$-6 \leq \alpha_Y,~\alpha_{\rm B-L},~\alpha_{\rm S} \leq 6$ 
has been calculated in every $0.2$ intervals for 
$\alpha_Y$ and $\alpha_{\rm B-L}$. 
We search the allowed region for which the correct electroweak symmetry breaking
and positive slepton mass squared are realized.
In Fig.~\ref{Fig1}, we present the allowed parameter sets of 
 ($\alpha_Y, \alpha_{\rm B-L})$ for fixed $\alpha_{\rm S}=0,\;1,\,2,\,4$. 
It is shown that the effect of adding the U(1)$_S$ D-term can expand the range
of allowed parameter spaces.
\footnote{As shown in Ref.~\cite{Martin}, the allowed region can change
a bit from Eq.~\bref{ineq} due to the RG effect of $S={\rm Tr}[Y  m^2]$.}

In Fig.~\ref{Fig2} and Fig.~\ref{Fig3}, we show the D-term dependence of the sparticle masses. 
We present mass spectra at some points in Table~\ref{table1}. 
In Fig.~\ref{Fig2} and Fig.~\ref{Fig3}, it is shown the spectra as a function of $\alpha_{\rm S}$ and $\alpha_Y$
with a fixed set of D-terms as $(\alpha_Y,\,\alpha_{\rm B-L})=(0,\,0)$ for Fig.~\ref{Fig2} and
as $(\alpha_{\rm B-L},\, \alpha_S)=(0,\,5)$ for Fig.~\ref{Fig3}, respectively.
In the plotted region, all the sfermion squared masses are positive 
and it realizes the correct electroweak symmetry breaking.
%
There exist some characteristic features of the D-term assisted AMSB scenario.
One important fact is that as is shown in Fig.~\ref{Fig2} one can solve 
the tachyonic slepton problem by adding only one D-term, $D_{\rm S}$,
which has been motivated by considering the E$_6$ models.
Another important effects of the D-term is the mass splitting between the left-handed
and right-handed sfermions.
%
%
Specifically, the left-handed stau can become lighter than the right-handed stau, 
which is unusual because, in general, the RG running due to the SU(2) gauge interaction 
pushes up the left-handed slepton mass heavier than the right handed one.
Therefore, it is really interesting to see the sfermion mass splittings as a distinct signature
of this scenario at the LHC.

The gaugino sector is the same as in the pure AMSB case. 
The mass ratios are approximately $M_1 : M_2 : M_3 = 3 : 1 : 7$.
So the lightest SUSY particle (LSP) is the Wino (rather than the more conventional LSP, Bino).
Those predictions for the gaugino masses in the AMSB has interesting phenomenological consequences. 
The remarkable fact is that the lightest chargino mass is nearly degenerated 
with the lightest neutralino mass.

For the decay of squarks and sleptons, naively speaking, initially produced left-handed squark
mainly decays into the lightest neutralino ($\widetilde{\chi}^0_1$) or the lightest chargino
($\widetilde{\chi}^{\pm}_1$), and successively $\widetilde{\chi}^{\pm}_1$ decays into
a very soft charged pion ($\pi^\pm$) and $\widetilde{\chi}^0_1$:
\be
\widetilde{q}_L \to \widetilde{\chi}^0_1 + q\;, 
\label{decay1}
\ee
or
\be
\widetilde{q}_L \to \widetilde{\chi}^{\pm}_1 + q \to \pi^{\pm} + \widetilde{\chi}^0_1 + q \;.
\label{decay2}
\ee
On the contrary, right-handed squark mainly decays into the second lightest neutralino
($\widetilde{\chi}^0_2$) or the gluino, and successively $\widetilde{\chi}^0_2$ decays into
slepton-lepton pair, and the gluino decays into the lightest stop-top pair:
\be
\widetilde{q}_R \to \widetilde{\chi}^0_2 + q \to \widetilde{\ell}^\pm + \ell^\mp + q \;,
\label{decay3}
\ee
or
\be
\widetilde{q}_R \to \widetilde{g} + q  \to \widetilde{\overline{t}}_1 + t  + q~~(\mbox{or}~~\widetilde{t}_1 + \overline{t}  + q) \;.
\label{decay4}
\ee
From those cascade decays Eqs.~\bref{decay1}-\bref{decay4},
one can expect fewer jets with no lepton event for left-handed squark decay as is shown in Eqs.~\bref{decay1}-\bref{decay2}. 
On the other hand, one would expect to see jet or lepton multiplicity for right-handed squark decay as shown in Eqs.~\bref{decay3}-\bref{decay4}.

\subsection{Dark matter relic density}
In this section we discuss the cosmological features of the lightest neutralino.
The recent Wilkinson Microwave Anisotropy Probe (WMAP) satellite data \cite{WMAP5}  provide
estimations of various cosmological parameters with greater accuracy. 
The current density of the universe is composed of about 73\% of dark energy and
27\% of matter. Most of the matter density is in the form of the CDM, 
and its density is 
estimated to be \cite{WMAP5} 
\begin{eqnarray}
\Omega_{\rm CDM} h^2  = 0.1143 \pm 0.0034  \;. 
 \label{WMAP} 
\end{eqnarray}
If the R-parity is conserved in SUSY models, the LSP is stable. 
The lightest neutralino, if it is the LSP, is the plausible candidate for the CDM. 

\begin{figure}[p]
\begin{center}
\includegraphics[width=.8\linewidth]{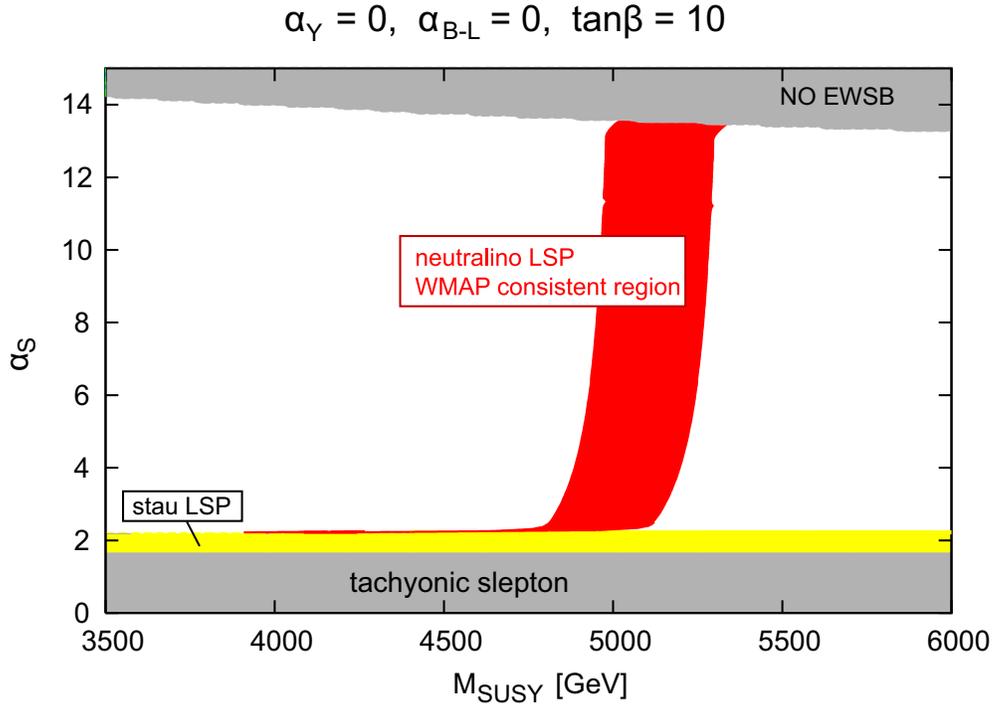}
\end{center}
\caption{
The red colored region (vertical band) corresponds to the region, in which the relic abundance
becomes consistent with the WMAP data: $0.1075 < \Omega_{\widetilde{\chi}_1^0} h^2 < 0.1211$.
This region also provides all the sfermion squared masses positive and the correct electroweak symmetry breaking.
}
\label{Fig4}
\vspace{1cm}
\end{figure}

Now we evaluate the relic abundance of the neutralino DM in this model by using {\tt micrOMEGAs} \cite{Belanger:2006is}.
The similar study in the context of the minimal AMSB scenario has been carried out in \cite{Chattopadhyay:2006xb}.
In Fig.~\ref{Fig4}, we show the WMAP allowed region in the parameter space $(M_{\rm SUSY},\, \alpha_{\rm S})$.

In the AMSB scenario, the lightest neutralino is mostly Wino-like, 
and it undergoes rapid annihilation through the reaction:
$\widetilde{W} \widetilde{W} \to W^+ W^-$. 
The resultant relic abundance is too small, which can roughly be estimated to be \cite{AMSB2}
\be
\Omega_{\widetilde{W}} h^2 \simeq 5 \times10^{-4} \, 
\left(\frac{M_{\widetilde{W}}}{100~\mbox{GeV}} \right)^2 \; . 
\label{WMAP2}
\ee
So the mass of the DM neutralino has to be very heavy to satisfy the WMAP data.
In fact, the numerical result in Fig.~\ref{Fig4} 
shows $M_{\rm SUSY}$ has to be about $5$ TeV as explained in Eq.~\bref{WMAP2}.
The horizontal thin line in Fig.~\ref{Fig4} corresponds to the so called, 
stau co-annihilation region. 
If the Wino-like neutralino with SU(2)$_L$ charge is much heavier than the weak gauge boson as described above, 
the weak interaction is a long-distance force for non-relativistic two-bodies states of such particles. 
If this non-perturbative effect (namely, Sommerfeld enhancement) of the dark matter at the freeze-out
temperature is taken into account, the abundance can be reduced by about 50\% \cite{non-pert1, Hisano:2006nn}.
Therefore, the allowed region shifts toward larger value of $M_{\rm SUSY}$.

Such a large value of $M_{\rm SUSY}$ is disfavored in view of the little hierarchy problem.
In order to keep the neutralino DM light, non-thermal production of the DM should be considered
as proposed in \cite{Moroi-Randall}. 
Once we accept the non-thermal production of the LSP neutralino from the moduli decays,
then it is possible to produce sufficient relic abundance of the LSP neutralino
even for the light Wino-like neutralino DM.

%
%
%

\section{Summary and discussion}
Anomaly mediation of supersymmetry breaking (AMSB) is very attractive 
because the resultant soft supersymmetry breaking parameters 
at a given energy scale are determined only by physics at that energy scale
(UV insensitivity) and hence is highly predictive (only one parameter, $F_\phi$). 
However, there is tachyonic slepton problem.
It is known that adding some D-terms such as 
${\rm U}(1)_Y$ and ${\rm U}(1)_{\rm B-L}$ is the well-motivated solution to this problem.

In this paper, we have considered the effects of additional D-terms,
which might be originated from the E$_6$ GUT models.
We have evaluated the soft SUSY breaking terms and obtained various sparticle mass spectra 
 for various input values of ($\alpha_Y$, $\alpha_{\rm B-L}$, $\alpha_{\rm S}$), 
 that are different from those obtained in the conventional anomaly mediation. 
It has been found that even if we add only one extra ${\rm U}(1)_{\rm S}$ D-term 
(without U(1)$_Y$ and U(1)$_{B-L}$ D-terms), 
it can become a solution to the tachyonic slepton problem, so it is more economical.
Since there could have some amounts of mass splittings between the right-handed 
sfermion and the left-handed one, this scenario can have a very distinct signature at the LHC.

We have also evaluated the dark matter relic density in our scenario, and we have shown that 
there exist parameter space, which is consistent with the WMAP observational data. In the WMAP
consistent region, the mass of the thermal relic DM is required to be heavy since the lightest neutralino
is Wino-like. However, if one consider the non-thermal production of the neutralino DM, 
there is a possibility to keep the neutralino DM light.

\section*{Acknowledgments}
T.K. and S.-G.K. would like to thank K.S. Babu for his hospitality
at Oklahoma State University.
The work of T.K. is supported by the Research
Fellowship of the Japan Society for the Promotion of Science (\#1911329).
The work of S.-G.K. is supported by the Research
Fellowship of the Japan Society for the Promotion of Science (\#206630).
We also particularly thank Tim Jones for helpful correspondence.


\newpage
\pagestyle{empty}
%
%
\begin{table}[p]
\centering
\begin{tabular}{|c|c|c|c|c|c|}
\hline \hline
$\tan \beta$                    
                            & 10
                            & 10
                            & 10 \\
$M_{\rm SUSY}$ [GeV]                   
                            & 500
                            & 500
                            & 500 \\
$(\alpha_Y, \alpha_{\rm B-L}, \alpha_{\rm S})$  
                                & ($0$, $0$, $5$)
                                & ($-8$, $0$, $5$)
                                & ($2$, $0$, $5$) \\
\hline \hline
$m_{\widetilde{\chi}^0_{1,2,3,4}}$     
                         &  229.5, 718.8, 1604, 1606
                         &  229.0, 722.0, 1461, 1463
                         &  229.5, 718.4, 1638, 1640 \\
$m_{\widetilde{\chi}^{\pm}_{1,2}}$ 
                         & 229.6, 1609
                         & 229.2, 1466
                         & 229.7,  1643 \\
$m_{\widetilde{g}}$          
                         & 1669
                         & 1686
                         & 1689 \\
\hline
$m_{{\widetilde{e},\widetilde{\mu}}_{L,R}}$ 
                          & 863.8, 525.0
                          & 557.5, 1073
                          & 925.0, 236.6\\
$m_{\widetilde{\tau}_{1,2}}$  
                          & 517.6, 861.9
                          & 547.9, 1069
                          & 221.6, 923.4 \\
\hline
$m_{\widetilde{\nu}_{e,\mu}}$ 
                          & 858.9
                          & 544.2
                          & 920.8   \\
$m_{\widetilde{\nu}_{\tau}}$  
                          & 856.3
                          & 533.3
                          & 918.9   \\
\hline
$m_{{\widetilde{u},\widetilde{c}}_{L,R}}$  
                         & 1689, 1705   
                         & 1732, 1515   
                         & 1677, 1750    \\
$m_{\widetilde{t}_{1,2}}$    
                         & 1326, 1551
                         & 1072, 1577
                         & 1375, 1548 \\
\hline
$m_{{\widetilde{d},\widetilde{s}}_{L,R}}$  
                         & 1691, 1832
                         & 1734, 1911
                         & 1679, 1811 \\
$m_{\widetilde{b}_{1,2}}$    
                         & 1515, 1807   
                         & 1560, 1883   
                         & 1503, 1787    \\
\hline
$m_h$                    & 116.1
                         & 116.2
                         & 116.1 \\
$m_H$             
                         & 1095
                         & 542.1
                         & 1194 \\
$m_A$              
                         & 1087
                         & 538.2
                         & 1186 \\
$m_{H^{\pm}}$            
                         & 1097
                         & 547.5
                         & 1197 \\
\hline \hline
\end{tabular}
\caption{
Sparticle and Higgs boson mass spectra (in units of GeV) 
 in the case of $\tan \beta=10$ and $m_t=172.5 ~{\rm GeV}$.}
\label{table1}
\end{table}
%
\end{document}